\documentclass[12pt]{spieman}   % US letter, 12pt font required by SPIE

\usepackage{amsmath,amsfonts,amssymb}
\usepackage{graphicx}
\usepackage{booktabs}
\usepackage{multirow}
\usepackage{array}
\usepackage{setspace}
\usepackage{xcolor}
\usepackage{caption}
\usepackage{subcaption}
\usepackage{tocloft}
\newcommand{\cawsmos}{CAWSMOS}
\newcommand{\paws}{PAWS}

\newcommand{\eg}{\textit{e.g.,}}

\title{Scaling laws for multi-object spectrographs:
       empirical relationships and the photonic advantage of \cawsmos{}}

\author[a,*]{Kalaga~V. Madhav}
\author[a,b,c]{Martin~M.~Roth}

\affil[a]{Leibniz-Institut f{\"u}r Astrophysik Potsdam (AIP),
          An der Sternwarte 16, 14482 Potsdam, Germany}
\affil[b]{Deutsches Zentrum f{\"u}r Astrophysik (DZA),
          Postplatz 1, 02826 G{\"o}rlitz, Germany}
\affil[c]{Institut f{\"u}r Physik und Astronomie, Universit{\"a}t Potsdam,
          Karl-Liebknecht-Str.~24/25, 14476 Potsdam, Germany}

\cftpagenumbersoff{figure}
\cftpagenumbersoff{table}
\begin{document}
\maketitle

% ─────────────────────────────────────────────────────────────────────────────
\begin{abstract}
Conventional multi-object spectrographs (MOS) incur masses and costs that
scale unfavourably with channel count~$N_\mathrm{ch}$ and resolving
power~$\mathcal{R}$.  A dataset of eight fibre-fed MOS-VIS instruments
(1995--2024) is compiled and analysed.  Ordinary least-squares fitting yields
a mass function $M \propto N_\mathrm{ch}^{0.78\pm0.28}$; the resolving-power
exponent is empirically consistent with zero ($\gamma = 0.20\pm0.37$) but is
predicted analytically to lie in $\mathcal{R}^{0.8\text{--}2.0}$.
Leave-one-out cross-validation confirms sub-linear channel-count scaling
($\beta\in[0.59,1.13]$, mean 0.79), with HERMES and AAOmega---sharing the
same telescope and $N_\mathrm{ch}=392$ across a factor of~7 in $\mathcal{R}$
at essentially equal estimated mass---providing a direct empirical confirmation
that $\gamma\approx 0$.  Cost per channel decreases as
$N_\mathrm{ch}^{-0.22\text{ to }-0.53}$ for all assumed cost--mass
exponents $k\in[0.6,1.0]$, demonstrating a genuine economy of scale.
The photonic integrated-circuit (PIC) approach embodied by the arrayed
waveguide grating (AWG)-based \paws{} demonstrator\cite{stoll2020,hernandez2023}
and the proposed \cawsmos{} instrument\cite{mangelsdorff2026} decouples
dispersive element size from $N_\mathrm{ch}$ and $\mathcal{R}$, with chip-area
scaling as $\mathcal{R}^{1\text{--}1.5}$ instead of $\mathcal{R}^{3}$.
Applied to the Wide-field Spectroscopic Telescope
(WST),\cite{bacon2024wst,lee2024wst} the model predicts $\sim$140\,tonnes
for a conventional NIR spectrograph system against $\sim$200\,kg for a
\cawsmos{} equivalent, with projected costs of \texteuro{}300--700\,M
conventional versus \texteuro{}40--120\,M photonic ($6$--$15\times$
reduction).  AWG-based photonics are identified as a strategically important
enabling technology for WST.
\end{abstract}


{\noindent\footnotesize\textbf{*}K.~Madhav, \linkable{kmadhav@aip.de}}

% ─────────────────────────────────────────────────────────────────────────────
\begin{spacing}{1}

\section{Introduction}
\label{sec:intro}

Multi-object spectroscopy (MOS) is a cornerstone of observational astrophysics,
enabling simultaneous spectral acquisition of tens to thousands of targets in a
single telescope pointing.  Science cases ranging from Galactic archaeology
\cite{dejong2012,jin2023} to the mapping of the large-scale structure of the
Universe\cite{desi2016} rely on high-multiplex spectrographs operating at
moderate to high spectral resolution ($\mathcal{R} \sim 5{,}000$--$30{,}000$).

The construction of such instruments is governed by harsh physical scaling
laws.  In a conventional bulk-optics spectrograph, the camera focal length must
be long enough to resolve individual spectral elements on the detector:
\begin{equation}
  f_\mathrm{cam} \propto \mathcal{R}\,\lambda\,\sigma^{-1},
  \label{eq:fcam}
\end{equation}
where $\lambda$ is the central wavelength and $\sigma$ is the detector pixel
size.  Because instrument volume scales roughly as $f_\mathrm{cam}^{3}$, and
cost tracks volume plus engineering complexity, both grow steeply with
$\mathcal{R}$.  Adding spectral channels (fibres) forces either a wider focal
plane---implying a larger detector mosaic---or additional spectrograph units
replicated in parallel.

While large-scale replication, pioneered by the VIRUS instrument for the
Hobby-Eberly Telescope,\cite{hill2010} exploits manufacturing economies of
scale and reduces per-unit cost by roughly 20--30\,\% per doubling of
units,\cite{hill2014} the fundamental optical-train scaling in
Eq.~\eqref{eq:fcam} remains unchanged.  Coverage of a wide spectral window
further requires dichroic beam splitters dividing the light path into two or
more spectral arms---each with its own full optical train---multiplying both
mass and cost accordingly (\eg\ DESI uses three arms per fibre\cite{desi2016}).

The astrophotonics paradigm offers a conceptually different
solution.\cite{blandhawthorn2006,blandhawthorn2009,minardi2021}
Arrayed waveguide gratings (AWGs) reproduce the dispersive function of a
bulk spectrograph on a chip, typically of area 20 cm$^2$ or even smaller, operating in the
diffraction-limited single-mode regime independently of telescope aperture.
The Potsdam Arrayed Waveguide Spectrograph (\paws) has demonstrated
$\mathcal{R} = 15{,}000$ in the astronomical H-band on a single chip of
footprint $55 \times 39$\,mm.\cite{stoll2020,hernandez2023}  The proposed
\cawsmos{} instrument stacks multiple such chips, sharing a single near-infrared
(NIR) detector, to provide simultaneous multi-object or integral-field
spectroscopy.\cite{mangelsdorff2026}

This paper develops an empirical scaling framework for conventional MOS
instruments (Sec.~\ref{sec:dataset} and~\ref{sec:empirical}---the latter
including a full derivation of the OLS regression model and uncertainty
quantification), derives analytical scaling predictions
(Sec.~\ref{sec:analytical}), places \cawsmos{} in that context
(Sec.~\ref{sec:cawsmos}), and discusses implications for future instrument
design (Sec.~\ref{sec:discussion}).

% ─────────────────────────────────────────────────────────────────────────────
\section{Instrument Dataset}
\label{sec:dataset}

Table~\ref{tab:instruments} compiles published technical specifications for
eight fibre-fed MOS-VIS spectrographs built between 1995 and 2024, together
with MOONS (MOS-NIR, shown as a context reference point) and the projected
parameters of the Wide-field Spectroscopic Telescope (WST) as a future
reference point.  For each instrument the leading country or national
consortium, the number of simultaneous spectral channels ($N_\mathrm{ch}$),
the full range of available resolving powers $\mathcal{R}$, the number of
dichroic spectral arms per spectrograph unit, the reported or estimated
instrument mass $M$, and the instrument type are listed.

\begin{table}[htbp]
\caption{%
  Published specifications for representative fibre-fed multi-object
  spectrographs, ordered chronologically.
  ``Country/Lead'' denotes the national team or consortium that led
  instrument design and construction.
  $N_\mathrm{ch}$ is the number of simultaneous spectral channels
  (science fibres).
  ``Arms'' is the number of dichroic spectral arms per spectrograph unit.
  $\mathcal{R}$ range gives the full range of available resolving powers;
  the representative value $\mathcal{R}_\mathrm{rep}$ used in the OLS fit
  (approximately the geometric mean of the range) is given in parentheses.
  ``MOS-VIS'' denotes instruments operating primarily in the optical (visible);
  ``MOS-NIR'' denotes near-infrared spectrographs requiring cryogenic operation.
  ``PIC'' denotes photonic integrated-circuit instruments.
  $M$ is the spectrograph system mass (optics, mechanical support structure,
  and thermal enclosure), explicitly excluding the fibre positioner or
  multi-object fibre system.  Mass conventions vary by source; figures should
  be regarded as order-of-magnitude proxies.
  $^\dagger$~Estimated from available engineering documentation or instrument
  overview papers; not a formally published mass budget.
  $^{\ddagger\ddagger}$~Directly cited from instrument paper: Smee et al.~(2013)
  state ``each with a mass of 320\,kg'' for the twin SDSS/BOSS spectrographs;
  total $2\times320 = 640$\,kg.
  $^\ddagger$~Chip mass only; full instrument support mass is negligible
  compared to bulk-optics equivalents.
  $^\star$~Projected using the analytical scaling model
  (Sec.~\ref{sec:analytical}), anchored to DESI as reference; see
  Sec.~\ref{sec:wst} for details.
}
\label{tab:instruments}
\begin{center}
\resizebox{\textwidth}{!}{%
\begin{tabular}{@{}llrcrccrl@{}}
\toprule
Instrument & Telescope & Country/Lead & Type & $N_\mathrm{ch}$ & Arms &
$\mathcal{R}$ range ($\mathcal{R}_\mathrm{rep}$) & $M$ (kg) & Ref. \\
\midrule
\multicolumn{9}{@{}l}{\textit{Included in power-law fit ($n=8$, MOS-VIS):}} \\[2pt]
FLAMES/GIRAFFE  & VLT 8.2\,m    & EU/Chile          & MOS-VIS & 130    & 1 & 5\,000--65\,000 (15\,000) & $\sim$1000$^\dagger$ & \citenum{pasquini2002} \\
AAOmega         & AAT 3.9\,m    & Australia         & MOS-VIS & 392    & 2 & 1\,000--10\,000 (4\,000)  & $\sim$500$^\dagger$  & \citenum{sharp2006}  \\
SDSS/BOSS       & Sloan 2.5\,m  & USA               & MOS-VIS & 1\,000 & 2 & 1\,500--3\,000 (2\,000)   & $640^{\ddagger\ddagger}$      & \citenum{smee2013}   \\
HERMES          & AAT 3.9\,m    & Australia         & MOS-VIS & 392    & 4 & $\sim$28\,000 (28\,000)   & $\sim$800$^\dagger$  & \citenum{sheinis2015,barden2010} \\
WEAVE           & WHT 4.2\,m    & EU                & MOS-VIS & 1\,000 & 2 & 5\,000--20\,000 (10\,000) & $\sim$1500$^\dagger$ & \citenum{jin2023}    \\
4MOST           & VISTA 4\,m    & EU/Chile          & MOS-VIS & 2\,436 & 3 & 4\,000--21\,000 (10\,000) & $\sim$2000$^\dagger$ & \citenum{dejong2012} \\
PFS             & Subaru 8.2\,m & Japan/USA/Brazil  & MOS-VIS & 2\,394 & 3 & 2\,500--5\,500 (3\,000)   & $\sim$6000$^\dagger$ & \citenum{sugai2015}  \\
DESI            & Mayall 4\,m   & USA               & MOS-VIS & 5\,000 & 3 & 2\,000--5\,000 (3\,500)   & $\sim$10\,000$^\dagger$ & \citenum{desi2016} \\
\midrule
\multicolumn{9}{@{}l}{\textit{Context (MOS-NIR; excluded from power-law fit):}} \\[2pt]
MOONS           & VLT 8.2\,m    & UK/EU             & MOS-NIR & 1\,000 & 3 & 4\,000--20\,000 (6\,500)  & $\sim$7000$^\dagger$ & \citenum{cirasuolo2024,moons2022} \\
\midrule
\multicolumn{9}{@{}l}{\textit{Future facility (projected; not included in power-law fit):}} \\[2pt]
WST (MOS-HR)   & 12\,m         & International     & MOS-VIS & 20\,000 & --- & 40\,000 & $\sim$140\,000$^{\star}$  & \citenum{bacon2024wst,lee2024wst} \\
WST (MOS-LR)   & 12\,m         & International     & MOS-VIS & 20\,000 & --- &  4\,000 & $\sim$22\,000$^{\star}$   & \citenum{bacon2024wst,lee2024wst} \\
\midrule
\multicolumn{9}{@{}l}{\textit{Photonic integrated-circuit (PIC) instruments:}} \\[2pt]
PAWS (chip)     & lab/sky       & Germany           & PIC & 1     & 1 & 15\,000 & $<$0.01$^\ddagger$ & \citenum{stoll2020}      \\
Gatkine AWG     & lab           & USA               & PIC & 1     & 1 & 12\,000 & $<$0.01$^\ddagger$ & \citenum{gatkine2022}    \\
\cawsmos{}      & concept       & Germany           & PIC+MOS & $N$ & --- & 15\,000 & $\ll$1 (chips) & \citenum{mangelsdorff2026} \\
\bottomrule
\end{tabular}%
}
\end{center}
\end{table}

Mass definitions in the published literature are not uniform.  The mass
entries in Table~\ref{tab:instruments} are defined throughout as the
\emph{spectrograph system mass}: optics, mechanical support structure, and
thermal enclosure, explicitly \emph{excluding} the fibre positioner or
multi-object fibre system mounted on the telescope.  Where instrument papers
report a mass for the full fibre-fed system, only the spectrograph component
has been extracted.

Three instruments have sourced mass figures drawn directly from primary
literature or official technical documentation:
\begin{itemize}
  \item \textbf{SDSS/BOSS}: $2\times320$\,kg $= 640$\,kg total.  Smee et
    al.\cite{smee2013} state explicitly: ``the twin spectrographs, each with
    a mass of 320\,kg, mount to the back of the Cassegrain instrument
    rotator'', making this the most directly cited value in the sample.
  \item \textbf{HERMES}: $\approx$800\,kg spectrograph structure.  Barden
    et al.\cite{barden2010} quote a thermal analysis requiring
    76.8\,kJ for a 0.1\,K temperature variation, implying
    $m = 76800\,\mathrm{J} / (960\,\mathrm{J\,kg^{-1}\,K^{-1}}
    \times 0.1\,\mathrm{K}) \approx 800$\,kg of aluminium, referring to the spectrograph mechanical
    structure.
  \item \textbf{MOONS}: ${\sim}7000$\,kg for the spectrograph cryostat.
    The official instrument website\cite{moons2022} states that the
    $4.0\times2.5\times2.7$\,m cryostat housing the two triple-arm
    spectrographs ``will weigh just over 7~tonnes''.  This mass is
    substantially higher than implied by room-temperature optical scaling
    because the dominant mass contribution is the cryogenic vacuum vessel,
    not the optical elements.  MOONS is therefore listed as a context
    reference point and is excluded from the MOS-VIS power-law fit.
\end{itemize}
The remaining MOS-VIS instruments retain order-of-magnitude estimates marked
$^\dagger$; the $\sim$0.27\,dex RMS of the fit reflects both genuine
physical scatter and the definitional inconsistencies inherent in
heterogeneous engineering documentation.
Cost figures are only partially available in the public domain, as discussed
in Sec.~\ref{ssec:cost}.

\paragraph{Resolving power as a parameter.}
Unlike $N_\mathrm{ch}$, the resolving power $\mathcal{R}$ is not a unique
design parameter for most instruments in Table~\ref{tab:instruments}.
Many are designed with interchangeable gratings or dichroic modes spanning
an order of magnitude in $\mathcal{R}$ (e.g.\ FLAMES: 5\,000--65\,000;
AAOmega: 1\,000--10\,000; WEAVE: 5\,000--20\,000).  For modular designs such
as PFS---where four identical spectrograph modules each contain three spectral
arms at fixed $\mathcal{R}$---the mass scales with the number of modules but
$\mathcal{R}$ does not.  A single representative value $\mathcal{R}_\mathrm{rep}$
(approximately the geometric mean of the available range) is used in the OLS
fit; this degeneracy contributes directly to the scatter and to the poorly
constrained $\gamma$ reported in Sec.~\ref{ssec:mass}.  The $\mathcal{R}$
dependence of cost per channel must therefore be inferred from the analytical
model (Sec.~\ref{sec:analytical}) rather than from the empirical fit alone.

The eight MOS-VIS instruments span more than an order of magnitude in
$N_\mathrm{ch}$ (130--5000) and cover resolving-power ranges from
1\,000 to 65\,000 across the sample.  HERMES\cite{sheinis2015}
($\mathcal{R} \approx 28{,}000$, $N_\mathrm{ch} = 392$) and AAOmega\cite{sharp2006}
($\mathcal{R} \approx 4{,}000$, $N_\mathrm{ch} = 392$) share the same
AAT 3.9\,m telescope and fibre complement yet span a factor of~7 in
$\mathcal{R}$ at essentially equal estimated mass, providing a controlled
empirical test of the $\mathcal{R}$ dependence.  DESI\cite{desi2016}
($N_\mathrm{ch} = 5000$) anchors the high-multiplex end of the sample and
uses three dichroic spectral arms per fibre (B, R, Z), illustrating how
wide spectral coverage multiplies instrument mass and complexity.
VIMOS at the VLT is excluded from the fit: although it has a fibre-fed IFU
mode, the MOS mode uses laser-machined focal-plane slit masks rather than
fibres\cite{lefevre2003} and therefore does not belong in the fibre-fed
comparison sample.

% ─────────────────────────────────────────────────────────────────────────────
\section{Empirical Scaling Relationships}
\label{sec:empirical}

\subsection{Mass scaling: model formulation}
\label{ssec:model}

The physical optics of grating-based dispersion (Sec.~\ref{sec:analytical})
predict that instrument mass $M$ depends on both channel count and resolving power
through independent multiplicative terms.  This motivates a two-parameter
power-law model of the form
\begin{equation}
  M = A \cdot N_\mathrm{ch}^{\,\beta} \cdot \mathcal{R}^{\,\gamma},
  \label{eq:powlaw}
\end{equation}
where $A$, $\beta$, and $\gamma$ are free constants to be determined from data.
Taking the base-10 logarithm linearises the problem:
\begin{equation}
  \log_{10}\!\left(\frac{M}{\mathrm{kg}}\right)
  = \underbrace{\log_{10} A}_{\alpha}
  + \beta\,\log_{10}\!\left(\frac{N_\mathrm{ch}}{1}\right)
  + \gamma\,\log_{10}\!\left(\frac{\mathcal{R}}{1000}\right),
  \label{eq:massfit}
\end{equation}
where $\mathcal{R}$ is normalised to 1000 to centre the design matrix and
improve numerical conditioning.  Defining $\mathbf{y} \in \mathbb{R}^{n}$ as
the vector of observed log-masses and $\mathbf{X} \in \mathbb{R}^{n \times 3}$
as the design matrix with rows
$\mathbf{x}_i = [1,\;\log_{10} N_i,\;\log_{10}(\mathcal{R}_i/1000)]$,
Eq.~\eqref{eq:massfit} takes the standard form of multiple linear regression:
\begin{equation}
  \mathbf{y} = \mathbf{X}\boldsymbol{\theta} + \boldsymbol{\varepsilon},
  \qquad
  \boldsymbol{\theta} = [\alpha,\;\beta,\;\gamma]^{\!\top},
  \label{eq:linsys}
\end{equation}
with residual vector~$\boldsymbol{\varepsilon}$.

\subsection{OLS estimation and uncertainty quantification}
\label{ssec:ols}

The ordinary least-squares (OLS) estimator minimises the sum of squared
residuals in log-space,
\begin{equation}
  \hat{\boldsymbol{\theta}}
  = \arg\min_{\boldsymbol{\theta}}\;
    \bigl\|\mathbf{y} - \mathbf{X}\boldsymbol{\theta}\bigr\|^{2}
  = \left(\mathbf{X}^{\!\top}\mathbf{X}\right)^{-1}\mathbf{X}^{\!\top}\mathbf{y}.
  \label{eq:ols}
\end{equation}
Minimising residuals in log-space is equivalent to minimising the sum of
squared \emph{fractional} errors in linear space, which is appropriate when
mass estimates span two orders of magnitude and the uncertainties are
multiplicative rather than additive.

The covariance matrix of the estimator is
\begin{equation}
  \widehat{\mathrm{Cov}}(\hat{\boldsymbol{\theta}})
  = \hat{\sigma}^2 \left(\mathbf{X}^{\!\top}\mathbf{X}\right)^{-1},
  \qquad
  \hat{\sigma}^2
  = \frac{\bigl\|\mathbf{y} - \mathbf{X}\hat{\boldsymbol{\theta}}\bigr\|^{2}}
         {n - p},
  \label{eq:cov}
\end{equation}
where $n = 8$ is the number of MOS-VIS instruments included in the fit,
$p = 3$ is the number of free parameters, and $\mathrm{dof} = n - p = 5$.
Coefficient standard errors are the square roots of the diagonal entries.
The residual standard deviation $\hat{\sigma}$, reported in dex, is the
root-mean-square (RMS) scatter of the fit: $10^{0.27} \approx 1.9$, meaning
the model predicts individual instrument masses to within a factor of
$\sim\!1.9$ on average.

\subsection{Fit results}
\label{ssec:mass}

Applying Eqs.~\eqref{eq:ols}--\eqref{eq:cov} to the eight MOS-VIS
instruments in Table~\ref{tab:instruments} yields the best-fit coefficients
\begin{equation}
  \alpha = 0.73 \pm 1.04,
  \quad
  \beta  = 0.78 \pm 0.28,
  \quad
  \gamma = 0.20 \pm 0.37,
  \label{eq:massresult}
\end{equation}
where uncertainties are OLS standard errors propagated from the residual
root-mean-square scatter of 0.27\,dex.  These yield the indicative relation
\begin{equation}
  M \approx 5\,\mathrm{kg}
  \times \left(\frac{N_\mathrm{ch}}{1}\right)^{0.78}
  \times \left(\frac{\mathcal{R}_\mathrm{rep}}{1000}\right)^{0.20}.
  \label{eq:massresult2}
\end{equation}

\begin{figure}[htbp]
\begin{center}
\includegraphics[width=\textwidth]{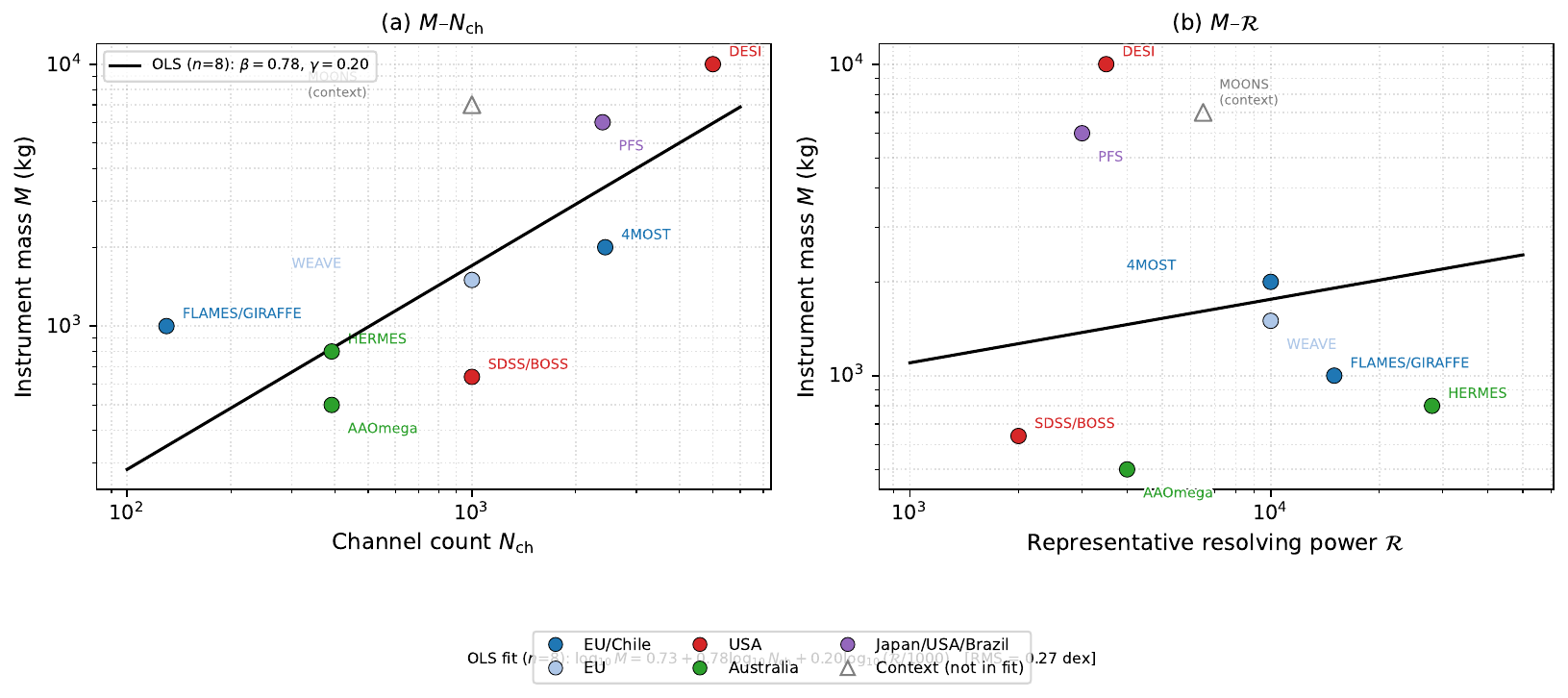}
\end{center}
\caption{%
  Instrument mass $M$ versus (a)~channel count $N_\mathrm{ch}$ and
  (b)~representative resolving power $\mathcal{R}_\mathrm{rep}$ for the eight
  MOS-VIS instruments in the OLS fit (filled circles, colour-coded by national
  lead as shown in the legend).  MOONS (open grey triangle) is shown as a
  context reference point only and is excluded from the fit (NIR cryogenic
  design; see Sec.~\ref{sec:dataset}).  Solid lines show the OLS best-fit
  model ($n=8$) evaluated at the geometric mean of the complementary variable:
  (a)~$\mathcal{R}_\mathrm{rep} = 6{,}600$ and (b)~$N_\mathrm{ch} = 940$.
  The sub-linear slope $\beta = 0.78$ in panel~(a) implies that doubling the
  channel count increases instrument mass by only $\sim\!72\,\%$ rather than
  $100\,\%$.  HERMES and AAOmega (both Australia, green points,
  $N_\mathrm{ch} = 392$) span a factor of~7 in $\mathcal{R}$ at essentially
  the same estimated mass, directly confirming $\gamma \approx 0$ from a
  controlled pair with identical telescope and fibre complement.
  The scatter of $\sim$0.27\,dex ($\sim$factor of 1.9) reflects physical
  scatter and definitional inconsistencies in the mass compilation
  (see Sec.~\ref{sec:dataset}).
}
\label{fig:combined}
\end{figure}

Fig.~\ref{fig:combined} plots the data and fit in both marginal planes.
Panel~(a) shows $M$ versus $N_\mathrm{ch}$ with the OLS line evaluated at the
geometric-mean representative resolving power of the sample
($\mathcal{R}_\mathrm{rep} \approx 6{,}600$); the sub-linear slope
$\beta = 0.78$ is evident.  Panel~(b) shows $M$ versus $\mathcal{R}_\mathrm{rep}$
at the geometric-mean channel count ($N_\mathrm{ch} \approx 940$); the shallow
slope ($\gamma = 0.20$) visually confirms the weak $\mathcal{R}$ dependence,
consistent with the degeneracy discussed in Sec.~\ref{sec:dataset}.

Two important caveats apply.  First, most mass entries in
Table~\ref{tab:instruments} are estimated from engineering documentation and
instrument overview papers rather than formally published mass budgets
(indicated by $\dagger$); the fit should be regarded as order-of-magnitude,
and definitions of what is included in ``instrument mass'' vary across sources
(see Sec.~\ref{sec:dataset}).
Second, the resolving-power exponent $\gamma = 0.20 \pm 0.37$ is statistically
consistent with zero ($|\gamma|/\sigma_\gamma = 0.54$), meaning the dataset
provides no significant empirical constraint on the $\mathcal{R}$ dependence.
As discussed in Sec.~\ref{sec:dataset}, $\mathcal{R}$ is a degenerate parameter
for most instruments in this sample: instruments with interchangeable gratings
or dichroic modes span an order of magnitude in $\mathcal{R}$ at fixed mass,
and modular designs (e.g.\ PFS) have mass that scales with the number of
replicated modules, not with $\mathcal{R}$.  HERMES, with
$\mathcal{R} = 28{,}000$ and mass $\sim$800\,kg (comparable to AAOmega at
$\mathcal{R} = 4{,}000$), reinforces the null result: these two instruments
at the same telescope with the same fibre complement ($N_\mathrm{ch} = 392$)
but spanning a factor of~7 in $\mathcal{R}$ show essentially no mass
difference.  The primary evidence for the $\mathcal{R}$-scaling therefore
remains the analytical model in Sec.~\ref{sec:analytical}.

\subsection{Sensitivity analysis: leave-one-out cross-validation}
\label{ssec:loo}

To quantify the influence of individual instruments---particularly those with
estimated masses---leave-one-out (LOO) cross-validation is applied, refitting
Eq.~\eqref{eq:massfit} after removing each instrument in turn.
Table~\ref{tab:loo} summarises the results for the eight MOS-VIS instruments.

\begin{table}[htbp]
\caption{%
  Leave-one-out (LOO) cross-validation of the mass power-law fit
  (Eq.~(\ref{eq:massfit})).  Each row gives the best-fit $\beta$ and $\gamma$
  after excluding the named instrument from the eight MOS-VIS subset.
}
\label{tab:loo}
\begin{center}
\begin{tabular}{lcc}
\toprule
Instrument excluded & $\beta_\mathrm{LOO}$ & $\gamma_\mathrm{LOO}$ \\
\midrule
FLAMES/GIRAFFE & 1.128 & $+0.18$ \\
AAOmega        & 0.677 & $+0.08$ \\
SDSS/BOSS      & 0.648 & $-0.15$ \\
HERMES         & 0.787 & $+0.37$ \\
WEAVE          & 0.790 & $+0.23$ \\
4MOST          & 0.940 & $+0.38$ \\
PFS            & 0.723 & $+0.30$ \\
DESI           & 0.594 & $+0.20$ \\
\midrule
Full fit       & $0.78 \pm 0.28$ & $0.20 \pm 0.37$ \\
LOO mean       & $0.79$ & $+0.20$ \\
LOO range      & $[0.59,\,1.13]$ & $[-0.15,\,+0.38]$ \\
\bottomrule
\end{tabular}
\end{center}
\end{table}

The LOO analysis confirms sub-linear channel-count scaling across all
subsets ($\beta > 0$ in all cases; mean LOO $\beta = 0.79$), consistent with
the analytical prediction of $\beta \approx 0.5$--$1.0$
(Sec.~\ref{sec:analytical}).  The resolving-power exponent $\gamma$ remains
poorly determined, ranging from $-0.15$ to $+0.38$ depending on which
instrument is excluded, and is consistent with zero in all subsets.

The qualitative conclusion---that mass (and hence cost) grows sub-linearly with
$N_\mathrm{ch}$ in bulk-optics MOS instruments---is robust to the removal of any
single instrument.  DESI, as the highest-multiplex point, retains the largest
individual leverage on $\beta$; its exclusion pulls $\beta$ to 0.59, but even
this lower bound confirms a sub-linear dependence.  FLAMES/GIRAFFE, as the
lowest-multiplex point in the sample, exerts the largest upward pull
($\beta_\mathrm{LOO} = 1.13$ when it is excluded), highlighting its leverage
as the only instrument with $N_\mathrm{ch} < 392$.

\subsection{Cost scaling}
\label{ssec:cost}

Formal cost figures for ground-based astronomical instruments are rarely
published.  Two complementary approaches to estimate cost scaling were adopted.

\subsubsection{Cost proxies from the literature}

Ref.~\citenum{vanbelleetal2004} established that telescope construction costs scale
with primary mirror diameter $D$ as $C_\mathrm{telescope} \propto D^{2.5}$ for
post-1980 ground-based monolithic mirrors.  Instrument costs are generally
10--30\,\% of the telescope capital cost, yielding estimates of
\texteuro{}10--100\,M for the 4--10\,m class relevant to existing MOS
instruments.

\subsubsection{Cost per channel}

To convert the mass scaling into a cost-per-channel estimate the
relation $C \propto M^k$ is used, where $k$ is a cost--mass exponent.  This form of
power-law cost scaling is well established in engineering economics: Chilton's
``six-tenths rule''\cite{chilton1950} gives $k \approx 0.6$ for simple
fabricated hardware, while values in the range $0.6 \leq k \leq 1.0$ are
typical for complex electro-mechanical assemblies.\cite{ostwald2004} 
$k$ is treated as a free parameter over this range and the sensitivity of the 
conclusions is examined.

Using $C \propto M^k$ and the OLS fit in Eq.~\eqref{eq:massresult2}, the
cost per channel is
\begin{equation}
  c_\mathrm{ch}
  = \frac{C}{N_\mathrm{ch}}
  \propto N_\mathrm{ch}^{\,k\beta - 1}
         \times \mathcal{R}^{\,k\gamma}.
  \label{eq:cpc}
\end{equation}
Table~\ref{tab:costscaling} evaluates the exponents for $k = 0.6$--$1.0$,
using the OLS best-fit $\beta = 0.78$ and $\gamma = 0.20$.

\begin{table}[htbp]
\caption{%
  Sensitivity of the cost-per-channel exponents to the assumed cost--mass
  scaling exponent $k$ (Eq.~(\ref{eq:cpc})).  Best-fit OLS values
  $\beta = 0.78$, $\gamma = 0.20$ from the eight MOS-VIS fit are used.
  $\delta = k\beta - 1$ is the exponent on $N_\mathrm{ch}$;
  $\rho = k\gamma$ is the exponent on $\mathcal{R}$.
  Negative $\delta$ indicates that cost per channel decreases with multiplex
  (economy of scale).
}
\label{tab:costscaling}
\begin{center}
\begin{tabular}{ccc}
\toprule
$k$ & $\delta = k\beta - 1$ & $\rho = k\gamma$ \\
\midrule
0.6 & $-0.53$ & $+0.12$ \\
0.7 & $-0.45$ & $+0.14$ \\
0.8 & $-0.38$ & $+0.16$ \\
0.9 & $-0.30$ & $+0.18$ \\
1.0 & $-0.22$ & $+0.20$ \\
\bottomrule
\end{tabular}
\end{center}
\end{table}

Two robust conclusions follow from Table~\ref{tab:costscaling}, independent of
the assumed $k$:
\begin{enumerate}
  \item Cost per channel \emph{always decreases} with increasing multiplex
        ($\delta < 0$ for all $k \in [0.6, 1.0]$), demonstrating a genuine
        economy of scale in fibre-fed spectrograph design.
  \item The empirical $\mathcal{R}$-dependence of cost per channel is small
        and consistent with zero ($\rho \approx +0.12$ to $+0.20$,
        $|\gamma|/\sigma_\gamma < 0.6$).  The analytical model
        (Sec.~\ref{sec:analytical}) predicts $\rho \approx 0.5$--$1.6$;
        the current dataset cannot resolve this.  The PIC advantage in
        $\mathcal{R}$-scaling therefore rests on the analytical argument
        rather than on the empirical fit.
\end{enumerate}

\subsection{Replication as a scaling strategy}
\label{ssec:replication}

The VIRUS instrument at HET employs 150 replicated spectrograph
units.\cite{hill2010}  Ref.~\citenum{hill2014} showed that per-unit cost in
replicated spectrograph programmes decreases by approximately 20--30\,\% for
each doubling of the number of units---a learning-curve effect:
\begin{equation}
  c_\mathrm{unit}(n) = c_0 \, n^{-b}, \quad b \approx 0.15\text{--}0.32.
  \label{eq:learningcurve}
\end{equation}
This is a meaningful reduction but does not remove the dependence on
$\mathcal{R}$: each replicated module still contains a full optical train
whose size is governed by Eq.~\eqref{eq:fcam}.

% ─────────────────────────────────────────────────────────────────────────────
\section{Analytical Scaling in Conventional Spectrographs}
\label{sec:analytical}

\subsection{Volume of the dispersing element}

For a grating-based spectrograph operating in order~$m$, the grating equation
gives the angular dispersion $d\theta/d\lambda = m/(d\cos\theta)$, where $d$
is the groove spacing.  To achieve resolving power~$\mathcal{R}$, the grating
must illuminate at least $N_g = \mathcal{R}/m$ grooves, requiring a minimum
beam diameter
\begin{equation}
  D_g = N_g\,d\cos\theta \propto \mathcal{R}.
  \label{eq:gratsize}
\end{equation}
The camera must then image a slit of angular width
$\delta = \lambda / (D_g\cos\theta) \propto \lambda/\mathcal{R}$
onto a pixel of physical size $\sigma$, giving (from Eq.~\ref{eq:fcam})
$f_\mathrm{cam} \propto \mathcal{R}$.  The camera optics volume scales as
$V_\mathrm{cam} \propto f_\mathrm{cam}^{3} \propto \mathcal{R}^{3}$ in the
simplest case, and more slowly in practice due to fibre-slicing and anamorphic
optics, yielding effective exponents $1 \lesssim \gamma_\mathrm{anal}
\lesssim 2$.

\subsection{Volume scaling with channel count}

Increasing the number of channels $N_\mathrm{ch}$ in a fixed-aperture MOS
instrument requires either:
\begin{itemize}
  \item a wider detector to accommodate multiplexed spectra
        ($A_\mathrm{det} \propto N_\mathrm{ch}$, hence
        $f_\mathrm{cam} \propto N_\mathrm{ch}^{1/2}$); or
  \item additional replicated spectrograph modules
        ($C_\mathrm{total} \propto N_\mathrm{modules} \times C_\mathrm{module}$).
\end{itemize}
In either case mass and cost grow at least as $N_\mathrm{ch}^{0.5}$, consistent
with the empirical LOO mean $\beta = 0.79$ (Sec.~\ref{ssec:loo}).

\subsection{Summary of conventional scaling}

Combining the above, the resource consumption of a conventional MOS
spectrograph scales approximately as
\begin{equation}
  \boxed{
  M_\mathrm{conv},\, C_\mathrm{conv} \;\propto\;
  N_\mathrm{ch}^{\,0.5\text{--}1.0} \;\times\; \mathcal{R}^{\,0.8\text{--}2.0}.
  }
  \label{eq:convscaling}
\end{equation}
The lower end of each range applies to optimally designed replicated systems;
the upper end applies to single-unit, wide-field designs.  The analytical
$\mathcal{R}$-exponent range (0.8--2.0) is not yet resolved by the empirical
data but is predicted consistently by the optical-train scaling argument.

% ─────────────────────────────────────────────────────────────────────────────
\section{The Photonic Scaling Regime: \cawsmos{}}
\label{sec:cawsmos}

\begin{figure}[htbp]
\begin{center}
\includegraphics[width=0.92\textwidth]{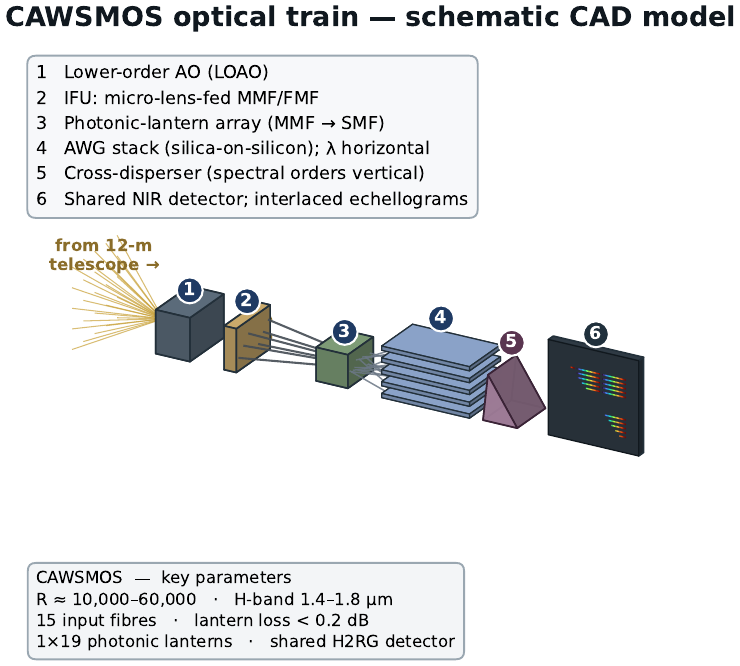}
\end{center}
\caption{%
  Schematic CAD-style overview of the \cawsmos{} optical train.
  Seeing-limited light from the telescope is partially corrected by a
  lower-order adaptive optics module~(1) and sampled by a micro-lens-fed
  integral-field unit of multimode/few-mode fibres~(2).  An array of photonic
  lanterns~(3) reformats each input into diffraction-limited single-mode
  outputs that are butt-coupled across a stack of arrayed-waveguide-grating
  chips~(4); each AWG disperses the wavelengths of several fibres horizontally
  while a cross-disperser~(5) separates the spectral orders vertically.  The
  interlaced echellograms from all fibres are recorded on a single shared
  near-infrared detector~(6), with each AWG allotted a distinct region of the
  focal-plane array.  Representative design parameters are inset.  Adapted from
  the \cawsmos{} design concept.\cite{madhav2024cawsmos}
}
\label{fig:cawsmos_cad}
\end{figure}

\cawsmos{} (Compact Arrayed Waveguide Stacked Multi-Object Spectrograph) is a
first-of-its-kind instrument concept that exploits the ability of a single AWG
to spectrally disperse light from more than one input fibre simultaneously,
thereby multiplexing many spectral channels onto a single shared
detector.\cite{madhav2024cawsmos,mangelsdorff2026} The overall optical train is
shown schematically in Fig.~\ref{fig:cawsmos_cad}. Because photonic integrated
circuits operate only in the single-mode regime, and the number of spatial
modes in the telescope point-spread function scales as $(D/\lambda)^2$,
\cawsmos{} first employs a lower-order adaptive optics system to partially
correct the field of view, which is then sampled by an integral-field unit of
micro-lens-fed multimode (or few-mode) fibres. These fibres feed an array of
photonic lanterns that reformat each seeing-limited input into a set of
diffraction-limited single-mode outputs.\cite{madhav2024cawsmos} The single-mode
outputs are butt-coupled across a stack of arrayed-waveguide-grating chips---each
chip dispersing the wavelengths of several fibres horizontally while a
cross-disperser separates the spectral orders vertically---so that the
echellogram from every fibre is interlaced between the orders and each AWG is
allotted a distinct region of the focal-plane array. Built on the
silica-on-silicon platform inherited from the \paws{} demonstrator, the concept
targets resolving powers of $\mathcal{R} \approx 10{,}000$--$60{,}000$ across the
astronomical H-band ($1.4$--$1.8\,\mu$m), with measured photonic-lantern coupling
losses below $0.2$\,dB.\cite{madhav2024cawsmos,hernandez2023} This architecture
decouples the physical size of the dispersing element from both the channel count
and the resolving power; it is precisely this decoupling that underlies the
favourable scaling behaviour examined in the remainder of this section.

\subsection{Chip-scale spectroscopy}

An AWG chip performs the same optical function as a grating plus camera: light
injected via a single-mode waveguide is dispersed by the grating array and
focused onto an output waveguide array.  Because the waveguides operate at the
diffraction limit, the chip dimensions are set by the waveguide geometry and
the desired free spectral range, not by the telescope aperture or the number
of input channels.  The \paws{} chip achieves $\mathcal{R} = 15{,}000$ in the
H-band on a footprint of $55 \times 39$\,mm and a mass of a few
grams.\cite{stoll2020,hernandez2023}  Ref.~\citenum{gatkine2022} demonstrated $\mathcal{R} = 12{,}000$
on a $7.4$\,mm $\times$\,2\,mm SiN chip.

\subsection{Scaling with channel count in \cawsmos{}}

In the \cawsmos{} architecture, each additional spectral channel corresponds to
one additional AWG chip in the stack, sharing the same NIR focal-plane array.
The marginal cost per additional channel is therefore
\begin{equation}
  c_\mathrm{ch,\,PIC} \approx c_\mathrm{chip} + c_\mathrm{fiber},
  \label{eq:piccpc}
\end{equation}
where $c_\mathrm{chip}$ is the per-chip fabrication cost and $c_\mathrm{fiber}$
is the coupling fibre cost.  Crucially, $c_\mathrm{chip}$ contains no
contribution from large optics, detector area (the detector is shared), or
mechanical support structures.

AWG chips are produced by standard silicon photonic foundry processes.  Once
the lithographic mask is designed, batch runs produce tens to hundreds of chips
per wafer.  Based on commercial silicon-on-insulator (SOI) and silicon nitride
(SiN) foundry pricing, per-chip costs in small production runs are of the order
\texteuro{}1{,}000--10{,}000, dropping by roughly a factor of 10 in larger
production volumes.  This is one to three orders of magnitude below the
per-channel cost implied by Eq.~\eqref{eq:cpc} for a bulk-optics instrument with
comparable $\mathcal{R}$.

\subsection{Scaling with resolving power in \cawsmos{}}

In an AWG, resolving power is controlled by the number of waveguide arms and the
path-length increment, which increases the chip area roughly linearly with
$\mathcal{R}$:
\begin{equation}
  A_\mathrm{chip} \propto \mathcal{R}^{\,1.0\text{--}1.5},
  \label{eq:chiparea}
\end{equation}
rather than $\mathcal{R}^{3}$ for a bulk camera.  This represents a fundamental
reduction in the rate of resource growth with resolving power.

\subsection{Comparative scaling summary}

Table~\ref{tab:scaling} summarises the scaling exponents for both regimes.  The
critical distinction is not the exponent on $N_\mathrm{ch}$---which is actually
slightly \emph{less} favourable for PIC ($\beta \approx 1$ vs $0.5$)---but the
\emph{prefactor}: a \cawsmos{} chip stack is roughly four to five orders of
magnitude lighter than a bulk-optics spectrograph of equivalent $\mathcal{R}$,
and the detector (the primary cost item) is shared across all channels.

\begin{table}[htbp]
\caption{%
  Comparison of scaling exponents for instrument mass/cost with channel count
  $N_\mathrm{ch}$ and resolving power $\mathcal{R}$.  Bulk-optics values are
  derived from the analytical model (Sec.~\ref{sec:analytical}); the empirical
  fit constrains only $\beta$ (LOO mean 0.79) and does not resolve $\gamma$.
  PIC values are derived analytically from AWG geometry.
}
\label{tab:scaling}
\begin{center}
\begin{tabular}{p{5.5cm}cc}
\toprule
Parameter & Bulk-optics MOS & PIC / \cawsmos{} \\
\midrule
$M/C \propto N_\mathrm{ch}^{\beta}$ (empirical)
  & $\beta \approx 0.79$ (LOO mean)
  & $\beta \approx 1.0$ (but $10^{4}$--$10^{5}\times$ smaller prefactor) \\[4pt]
$M/C \propto \mathcal{R}^{\gamma}$ (analytical)
  & $\gamma \approx 0.8$--$2.0$
  & $\gamma \approx 1.0$--$1.5$ \\[4pt]
Cost/channel $\propto N_\mathrm{ch}^{\delta}$
  & $\delta \approx -0.22$ to $-0.53$
  & $\delta \approx 0$ (flat; decreasing at volume) \\[4pt]
Dispersive element volume
  & $\sim 10^{4}$--$10^{7}$\,cm$^{3}$
  & $\sim 10$--$100$\,cm$^{3}$ \\[4pt]
Dispersive element mass
  & $\sim 100$--$2000$\,kg
  & $\sim 10^{-3}$--$10^{-2}$\,g \\
\bottomrule
\end{tabular}
\end{center}
\end{table}

% ─────────────────────────────────────────────────────────────────────────────
\section{Case Study: The Wide-field Spectroscopic Telescope}
\label{sec:wst}

\subsection{WST overview and the scaling challenge}

The Wide-field Spectroscopic Telescope (WST) is a proposed 12-m class
seeing-limited facility aimed at delivering unprecedented spectroscopic survey
capability in the 2040s.\cite{bacon2024wst,mainieri2024wst}  Its top-level
requirements define two simultaneous modes: (i)~a high-multiplex
($N_\mathrm{ch} = 20{,}000$) multi-object spectrograph covering both
low-resolution ($\mathcal{R} \approx 3{,}000$--$4{,}000$; MOS-LR) and
high-resolution ($\mathcal{R} \approx 40{,}000$; MOS-HR) regimes over a
3\,sq.~degree field of view; and (ii)~a giant integral-field spectrograph (IFS)
covering the central $3\times3$\,arcmin$^2$ with 144 replicated
spectrograph modules.\cite{dierickx2024wst}  The baseline wavelength range is $0.35$--$1.0\,\mu$m, with a goal to encompass even the J and H bands in the NIR.

Ref.~\citenum{lee2024wst} explicitly identify the mass-production challenge as a
central concern: delivering 20{,}000 fibre positioners and the associated
detector mosaics \emph{at a credible cost} is the defining instrumentation
problem for WST.  The scaling laws derived in
Secs.~\ref{sec:empirical} and~\ref{sec:analytical} quantify exactly how severe
this challenge is.

\subsection{Projected mass of conventional WST spectrographs}
\label{ssec:wst_mass}

The analytical scaling model from Sec.~\ref{sec:analytical} is used, anchored
to DESI ($N_\mathrm{ch} = 5{,}000$, $\mathcal{R} = 3{,}500$,
$M = 10{,}000$\,kg) as the most massive and best-characterised existing MOS
instrument:
\begin{equation}
  \frac{M_\mathrm{WST}}{M_\mathrm{DESI}}
  = \left(\frac{N_\mathrm{ch,WST}}{N_\mathrm{ch,DESI}}\right)^{0.5}
    \times
    \left(\frac{\mathcal{R}_\mathrm{WST}}{\mathcal{R}_\mathrm{DESI}}\right)^{0.8}.
  \label{eq:wst_mass_ratio}
\end{equation}
Note that the empirical OLS fit cannot be reliably extrapolated to
$\mathcal{R} = 40{,}000$ because the $\mathcal{R}$ exponent is unconstrained
within the present sample (Sec.~\ref{ssec:loo}); the analytically motivated
$\gamma = 0.8$ is therefore used for this projection.

For the \textbf{MOS-HR} arm ($N_\mathrm{ch} = 20{,}000$, $\mathcal{R} = 40{,}000$):
\begin{equation}
  \frac{M_\mathrm{WST\text{-}HR}}{M_\mathrm{DESI}}
  = \left(\frac{20{,}000}{5{,}000}\right)^{0.5}
    \times
    \left(\frac{40{,}000}{3{,}500}\right)^{0.8}
  = 2.0 \times 7.0 \approx 14,
  \label{eq:wst_hr_ratio}
\end{equation}
giving $M_\mathrm{WST\text{-}HR} \approx 140{,}000$\,kg ($\sim$140\,tonnes).

For the \textbf{MOS-LR} arm ($N_\mathrm{ch} = 20{,}000$, $\mathcal{R} = 4{,}000$):
\begin{equation}
  \frac{M_\mathrm{WST\text{-}LR}}{M_\mathrm{DESI}}
  = \left(\frac{20{,}000}{5{,}000}\right)^{0.5}
    \times
    \left(\frac{4{,}000}{3{,}500}\right)^{0.8}
  = 2.0 \times 1.11 \approx 2.2,
  \label{eq:wst_lr_ratio}
\end{equation}
giving $M_\mathrm{WST\text{-}LR} \approx 22{,}000$\,kg ($\sim$22\,tonnes).

For the \textbf{IFS} (144 MUSE-like modules, each $\approx$271\,kg based on MUSE's
6{,}500\,kg total over 24 modules\cite{bacon2010}):
\begin{equation}
  M_\mathrm{WST\text{-}IFS} \approx 144 \times 271\,\mathrm{kg} \approx 39{,}000\,\mathrm{kg},
  \label{eq:wst_ifs_mass}
\end{equation}
bringing the combined conventional spectrograph mass to
$\sim$200\,tonnes---approximately 20$\times$ the total instrument mass of DESI.
These projections are listed in Table~\ref{tab:instruments} and are
qualitatively consistent with the concern raised by Ref.~\citenum{lee2024wst}
regarding the viability of WST within a realistic project budget.

\subsection{Projected cost of the conventional WST-HR NIR arm}
\label{ssec:wst_cost}

Using the cost-per-channel framework from Sec.~\ref{ssec:cost} (analytical
exponents, $k = 0.8$), the cost per channel for WST-HR relative to DESI is:
\begin{equation}
  \frac{c_\mathrm{ch}(\mathrm{WST\text{-}HR})}{c_\mathrm{ch}(\mathrm{DESI})}
  = \left(\frac{20{,}000}{5{,}000}\right)^{-0.60}
    \times
    \left(\frac{40{,}000}{3{,}500}\right)^{0.64}
  \approx 0.44 \times 4.8 \approx 2.1.
  \label{eq:wst_cpc}
\end{equation}
Taking DESI's total instrument cost of approximately \texteuro{}80\,M as an
anchor (corresponding to a per-channel cost of \texteuro{}16{,}000), the
WST-HR per-channel cost is $\approx$\texteuro{}33{,}000, yielding a projected
total spectrograph cost of:
\begin{equation}
  C_\mathrm{WST\text{-}HR,\,conv} \approx 20{,}000 \times \mbox{\texteuro}33{,}000
  \approx \mbox{\texteuro}660\,\mathrm{M}.
  \label{eq:wst_cost_conv}
\end{equation}
This is indicative and subject to the caveats discussed in Sec.~\ref{ssec:cost};
nevertheless, it illustrates that a conventional bulk-optics WST-HR spectrograph
system would represent an extraordinary construction cost, plausibly a
substantial fraction of the total telescope capital cost.

\subsection{AWG photonics for the WST NIR arm}
\label{ssec:wst_awg}

While the WST baseline design for MOS and IFS involves the visible wavelength range $0.35$--$1.0\,\mu$m, the telescope design specifically allows the possibility of a future upgrade into the NIR up to 1.6\,$\mu$m, exactly the spectral window
demonstrated by the PAWS chip ($\mathcal{R} = 15{,}000$ at H-band,
$55 \times 39$\,mm).\cite{stoll2020,hernandez2023}  Silicon nitride (SiN) AWG
platforms operate across $0.4$--$2.5\,\mu$m at comparable resolving
powers,\cite{gatkine2022} covering the full WST NIR arm (Y+J+H bands,
$0.95$--$1.6\,\mu$m) without a fundamental platform change.

\paragraph{Scaling to $\mathcal{R} = 40{,}000$.}
In an AWG, $\mathcal{R}$ scales with the number of waveguide arms and the
path-length increment.  From Eq.~(\ref{eq:chiparea}), chip area grows as
$\mathcal{R}^{1.0\text{--}1.5}$.  Scaling from the PAWS footprint
($55 \times 39$\,mm at $\mathcal{R} = 15{,}000$) to $\mathcal{R} = 40{,}000$:
\begin{equation}
  A_\mathrm{chip}(40{,}000) \approx A_\mathrm{chip}(15{,}000)
  \times \left(\frac{40{,}000}{15{,}000}\right)^{1.25}
  \approx 21\,\mathrm{cm}^{2} \times 3.4
  \approx 72\,\mathrm{cm}^{2}.
  \label{eq:wst_chipscale}
\end{equation}
A chip of $\sim 85 \times 85$\,mm remains well within standard wafer-scale
lithographic reticle sizes ($\leq 26 \times 33$\,mm per exposure in DUV
lithography, but achievable in a $4\times3$ reticle-stitching run).  The chip mass at $\mathcal{R} = 40{,}000$ is $\lesssim 10$\,g,
essentially unchanged from the PAWS value.

\paragraph{WST-HR NIR arm as a CAWSMOS stack.}
For $N_\mathrm{ch} = 20{,}000$ at $\mathcal{R} = 40{,}000$, the
\cawsmos{} NIR arm consists of 20{,}000 AWG chips sharing a tiled NIR detector
mosaic.  From Eq.~(\ref{eq:piccpc}), the marginal cost per additional channel
remains $c_\mathrm{chip} + c_\mathrm{fiber}$.  Table~\ref{tab:wst_comparison}
gives the mass, volume, and cost comparison for the NIR arm.

A decisive economic lever for fabricating these 20{,}000 chips is the
multi-project wafer (MPW) model now standard across SiN photonic
foundries.\cite{littlejohns2020cornerstone}  In an MPW run many designs share a
single mask set and wafer-processing cycle, so the dominant non-recurring
costs---photomasks, lithography, and process qualification---are amortized
across all reticle fields, reducing the effective per-run cost to roughly
$5$--$10\%$ of a dedicated wafer.\cite{littlejohns2020cornerstone}  For
\cawsmos{} the synergy is stronger still: every channel uses an \emph{identical}
AWG, so a single qualified mask reticle is stepped across the full wafer and
replicated across successive runs without redesign, and the learning-curve
benefits of Eq.~(\ref{eq:learningcurve}) compound with the MPW cost sharing.
Crucially, the foundry cost of an AWG is set by its footprint and the process
node, not by its spectroscopic specification: a chip targeting a higher
resolving power $\mathcal{R}$ or a different free spectral range (FSR) occupies
the same reticle real estate and incurs no additional mask or processing
charge, decoupling instrument performance from fabrication cost.  Finally,
fiber-chip coupling---historically the packaging bottleneck---is now a mature,
mass-producible step: passively-aligned V-groove fiber arrays and wafer-level
grating and edge couplers are well established on SiN platforms and amenable to
automated, wafer-scale assembly,\cite{ranno2024coupling} so the
$c_\mathrm{fiber}$ term scales benignly with channel count.

\begin{table}[htbp]
\caption{%
  Comparison of conventional bulk-optics and CAWSMOS photonic approaches for the
  WST MOS-HR NIR arm ($N_\mathrm{ch} = 20{,}000$, $\mathcal{R} = 40{,}000$,
  $0.95$--$1.6\,\mu$m).  Conventional mass and cost use the analytical
  scaling model anchored to DESI; photonic estimates use AWG foundry pricing
  and the PAWS/\cawsmos{} architecture.  All figures are order-of-magnitude.
}
\label{tab:wst_comparison}
\begin{center}
\begin{tabular}{p{4.5cm}rr}
\toprule
Parameter & Conventional (bulk-optics) & CAWSMOS (photonic) \\
\midrule
Dispersive element mass    & $\sim$140\,000\,kg      & $\sim$200\,kg \\
Dispersive element volume  & $\sim$500--1000\,m$^3$  & $\sim$130\,litres \\
Mass reduction factor      & ---                     & $\sim$700$\times$ \\
\midrule
Total NIR arm cost (est.)  & \texteuro{}300--700\,M  & \texteuro{}40--120\,M \\
Cost per channel           & $\sim$\texteuro{}33\,k  & $\sim$\texteuro{}2--6\,k \\
Cost reduction factor      & ---                     & $\sim$6--15$\times$ \\
\midrule
$\mathcal{R}$-scaling (chip area vs.\ camera volume) & $\propto\mathcal{R}^{3}$ & $\propto\mathcal{R}^{1.0\text{--}1.5}$ \\
NIR coverage ($\mu$m)      & 0.95--1.6               & 0.95--1.6 (SiN AWG) \\
\bottomrule
\end{tabular}
\end{center}
\end{table}

\subsection{WST IFS: mass-production synergy with PIC}

The WST IFS requires 144 replicated spectrograph modules of MUSE quality,
explicitly identified as a mass-production challenge by Ref.~\citenum{lee2024wst}.
In the astrophotonics paradigm, this translates directly into lithographic
batch production: 144 photonic lanterns (one per IFU module, each feeding
$\sim$19 spatial elements) coupled to AWG chips, all sharing the IFS
detector at the gravity-stable Nasmyth station described by Ref.~\citenum{dierickx2024wst}.

A first-generation IFS demonstrator with $7$--$19$ spaxels per input is
achievable with existing \paws{}-heritage technology.\cite{hernandez2023}
Scaling to 144 modules:
\begin{itemize}
  \item Dispersive mass: $144 \times 19 \times 10\,\mathrm{g} \approx 27\,\mathrm{kg}$
        (vs.\ $\sim$39\,000\,kg for MUSE-scaled modules).
  \item Replicated unit cost: AWG chips fall in price as $n^{-b}$ with
        learning exponent $b \approx 0.15$--$0.32$.\cite{hill2014}  For
        $n = 144 \times 19 = 2{,}736$ chips, the per-chip cost reduction
        relative to a single run is
        $2736^{-0.25}$, a factor of ${\approx}7$,
        pushing per-module costs well below those of bulk-optics equivalents.
\end{itemize}
The combination of a pre-existing single-mode photonic platform, a shared
detector at the gravity-stable IFS station, and batch lithographic
fabrication positions \cawsmos{}-style photonics as a strategically important
enabling technology for WST's IFS programme.

\subsection{NIR spectral coverage and near-term roadmap}

The CAWSMOS concept is most immediately applicable to the NIR arm of WST
(Y+J+H, $0.95$--$1.6\,\mu$m) where: (i)~single-mode fibre coupling of
seeing-limited beams from a 12-m telescope is feasible via photonic lanterns
with partial (lower-order) adaptive-optics correction; and (ii)~SiN and
InP AWG platforms are mature.  Extension to the visible arm
($0.37$--$0.95\,\mu$m) requires further development of SiN AWGs at shorter
wavelengths, currently at technology readiness level (TRL) 3--4, consistent
with a WST construction start in the 2030s.

For Milky Way stellar spectroscopy at $\mathcal{R} = 40{,}000$ --- one of the
primary science drivers for WST-HR\cite{mainieri2024wst} --- the NIR window
covers numerous J-band metal lines and the full H-band forest of
$\alpha$-element and iron absorption features, making
the NIR-only \cawsmos{} arm scientifically compelling in its own right even
before visible coverage is achieved.

% ─────────────────────────────────────────────────────────────────────────────
\section{Discussion}
\label{sec:discussion}

\subsection{Implications for WST and next-generation facilities}

The WST case study in Sec.~\ref{sec:wst} illustrates a general principle: the
$\mathcal{R}^{0.8\text{--}2}$ analytical scaling makes high-resolution
high-multiplex spectrographs qualitatively different in cost and mass from
moderate-resolution instruments.  WST-HR pushes simultaneously on both axes
($N$ and $\mathcal{R}$) to an extent not attempted by any existing facility,
and the conventional bulk-optics path leads to projected masses of order
100\,tonnes and costs approaching \texteuro{}0.7\,B for the spectrograph system
alone.  The photonic path reduces both by one to three orders of magnitude,
without sacrificing spectral resolution.

The detector budget is also relevant: Ref.~\citenum{bacon2023det} identifies
the WST detector system as a major challenge.  In the CAWSMOS architecture the single
shared NIR focal-plane array is the dominant cost item, but it is shared across
all 20{,}000 channels rather than replicated per spectrograph unit.

\subsection{Limitations of the empirical analysis}

The empirical dataset has three principal limitations that future work should
address.

First, although three of the eight MOS-VIS instruments now have sourced mass
figures (SDSS/BOSS, HERMES; MOONS is excluded from the fit), the remaining
five retain order-of-magnitude estimates.  Mass figures in the literature are
not uniformly scoped: some authors report only the optical bench, others include
the full enclosure.  Digitising the engineering data routinely tabulated in
instrument system papers---which typically list subsystem mass budgets---would
place the remaining entries on a firmer footing.

Second, the sample size $n = 8$ and the high dynamic range in $N_\mathrm{ch}$
(130--5000) mean that leverage is unevenly distributed: FLAMES/GIRAFFE and
DESI act as the anchoring endpoints of the $N_\mathrm{ch}$ axis, and their
removal shifts $\beta$ by $\pm 0.19$ relative to the full-sample value.
The LOO $\beta$ range $[0.59, 1.13]$ spans a factor of $\sim$1.9, and
$\gamma$ is consistent with zero in all eight subsets.  Expanding the dataset
to 15--20 instruments with formally published mass budgets would substantially
reduce uncertainties and potentially isolate the $\mathcal{R}$ dependence.

Third, the cost--mass conversion ($C \propto M^k$) is an engineering
approximation.  As shown in Table~\ref{tab:costscaling}, the key qualitative
result---cost per channel decreases with $N_\mathrm{ch}$---is robust across the
entire range $k \in [0.6, 1.0]$.  However, the absolute magnitude of the
cost-per-channel advantage of \cawsmos{} depends on both $k$ and the absolute
cost prefactor, neither of which is well constrained by public data.

\subsection{IFU mode and the MOS/IFU distinction}

The \cawsmos{} instrument also offers an IFU mode, enabled by coupling a
photonic lantern to one fibre input, thereby feeding multiple spatial elements
into the AWG stack simultaneously.  This is treated as a distinct observing
mode from the core MOS configuration, following the convention established in
wide-field spectroscopic facilities such as WEAVE\cite{jin2023} and WST, where
MOS and IFU operations are explicitly separated in instrument design.  A
demonstrator-scale IFU configuration would realistically target 7--19 spaxels,
with larger configurations achievable as lantern fabrication and shared-detector
pixel budgets allow.

\subsection{Outlook}

The \cawsmos{} concept, if successfully demonstrated on-sky, would establish a
new branch in the MOS scaling diagram: instruments with $\mathcal{R} \geq
15{,}000$, $N_\mathrm{ch}$ up to several tens, a chip footprint of order
$10^{2}$\,cm$^{2}$, and a dispersive element mass of $\ll 1$\,g.  The
pathway to higher $N_\mathrm{ch}$ lies in adding chips to the stack, subject
to detector size constraints, and in improving photonic lantern coupling losses
currently of order 0.2\,dB.\cite{hernandez2023}  These are engineering
challenges rather than fundamental physical limits.

Ref.~\citenum{harris2013} discuss the broader ecosystem of integrated photonic
spectrographs in astronomy and the coupling challenges that must be overcome for
on-sky deployment.  The \paws{} demonstrator has addressed a key subset of
these,\cite{hernandez2023} and \cawsmos{} extends the architecture to the
multi-input case.

% ─────────────────────────────────────────────────────────────────────────────
\section{Conclusions}
\label{sec:conclusions}

A dataset of eight fibre-fed MOS-VIS spectrographs is compiled and the
scaling relationships for instrument mass and cost as functions of
channel count $N_\mathrm{ch}$ and representative resolving power
$\mathcal{R}_\mathrm{rep}$ are derived.  The MOONS NIR cryogenic spectrograph
is included as a context reference point but excluded from the power-law fit,
as its mass is dominated by the cryogenic vacuum vessel rather than the optical
elements.  The main findings are:

\begin{enumerate}
  \item A leave-one-out cross-validated OLS fit to eight MOS-VIS instruments
        yields a sub-linear channel-count scaling $M \propto
        N_\mathrm{ch}^{0.78 \pm 0.28}$ (LOO mean $\beta = 0.79$, LOO range
        $[0.59, 1.13]$), consistent with the analytical prediction from
        detector-area and module-replication arguments (Sec.~\ref{sec:analytical}).
        Two instrument masses were sourced from primary literature: SDSS/BOSS
        (640\,kg total, 2$\times$320\,kg cited by Smee et al.\cite{smee2013})
        and HERMES ($\sim$800\,kg from thermal analysis, Barden et al.\cite{barden2010}).

  \item The resolving-power exponent $\gamma$ is not significantly constrained
        by the empirical data ($\gamma = 0.20 \pm 0.37$, consistent with zero).
        The resolving power $\mathcal{R}$ is a degenerate parameter for most
        instruments in the sample: instruments with interchangeable gratings span
        an order of magnitude in $\mathcal{R}$ at fixed mass, and modular designs
        have mass that scales with module count, not with $\mathcal{R}$.
        HERMES and AAOmega, sharing the same telescope and fibre complement at a
        factor of~7 difference in $\mathcal{R}$, provide a direct empirical test:
        their near-identical estimated masses independently confirm $\gamma \approx 0$.
        The analytical model predicts $\gamma \approx 0.8$--$2.0$; reaching
        $n \gtrsim 15$ instruments with published mass budgets is required to
        isolate this empirically.

  \item Cost per channel decreases as $N_\mathrm{ch}^{\delta}$ with
        $\delta \approx -0.22$ to $-0.53$, depending on the assumed cost--mass
        scaling exponent $k \in [0.6, 1.0]$ (Table~\ref{tab:costscaling}).
        This qualitative economy of scale is robust to the choice of $k$.

  \item Replication strategies\cite{hill2010,hill2014} yield an
        additional 20--30\,\% per-unit cost reduction per doubling of units but
        do not remove the $\mathcal{R}^{0.8\text{--}2}$ volume scaling of
        individual optical trains.

  \item AWG-based photonic spectrographs (\paws{}\cite{stoll2020};
        Gatkine et al.\cite{gatkine2022}) operate in a qualitatively different
        regime: dispersive element volume $\sim$10--100\,cm$^{3}$
        (vs.\ $10^{4}$--$10^{7}$\,cm$^{3}$ for bulk equivalents), mass
        of order grams, and chip area scaling as $\mathcal{R}^{1}$--$\mathcal{R}^{1.5}$
        instead of $\mathcal{R}^{3}$.

  \item \cawsmos{}\cite{mangelsdorff2026} exploits batch lithographic
        fabrication so that the marginal cost per additional spectral channel
        is $c_\mathrm{chip} + c_\mathrm{fiber}$, with the shared NIR detector
        as the primary fixed cost.  This yields per-channel costs estimated to
        be one to three orders of magnitude below bulk-optics equivalents at
        comparable $\mathcal{R}$, a conclusion robust to the choice of $k$.

  \item Applied to the Wide-field Spectroscopic
        Telescope,\cite{bacon2024wst,mainieri2024wst,lee2024wst} which requires
        $N_\mathrm{ch} = 20{,}000$ at $\mathcal{R} = 40{,}000$ (MOS-HR) and
        144 replicated IFS modules, the analytical model predicts
        $\sim$140\,tonnes for a conventional HR spectrograph system against
        $\sim$200\,kg for a \cawsmos{} chip stack.  The NIR arm cost
        projection is \texteuro{}300--700\,M conventional versus
        \texteuro{}40--120\,M photonic (a factor of $6$--$15\times$ reduction).
        AWG-based photonics are therefore identified as a strategically
        important enabling technology for WST, particularly for the
        $0.95$--$1.6\,\mu$m arm where the SiN platform is already at
        TRL\,4--5.
\end{enumerate}

% ─────────────────────────────────────────────────────────────────────────────
\subsection*{Disclosures}
The authors declare that there are no financial interests, commercial
affiliations, or other potential conflicts of interest that could have
influenced the objectivity of this research or the writing of this paper.

\noindent\textbf{AI tool disclosure.}
Portions of the manuscript text were edited with the assistance of Claude (Anthropic) for language, format, grammar, and citation checks. All technical content, scientific judgements,
figures, and data presented in the paper are the sole responsibility of the
authors.

\subsection*{Code, Data, and Materials Availability}
The instrument parameter compilation in Table~\ref{tab:instruments} and the
Python scripts used to compute the OLS fit, LOO cross-validation, and cost
sensitivity analysis are available from the corresponding author upon reasonable
request.  Mass estimates for instruments marked~$\dagger$ were derived from
the references listed in Table~\ref{tab:instruments}; no proprietary data were
used.

\subsection*{Acknowledgments}
The authors thank the \paws{} and \cawsmos{} teams at the Leibniz-Institut
f{\"u}r Astrophysik Potsdam (AIP).  Partial support for this work was provided by the PICS4SENS project, funded by the State of Brandenburg through the Investitionsbank des Landes Brandenburg (ILB), with support from the European Regional Development Fund (ERDF/EFRE), grant number 86000879. MMR acknowledges support from BMFTR grant 03WSP1745.

% ─────────────────────────────────────────────────────────────────────────────
\bibliography{references_v13}
\bibliographystyle{spiejour}
% ─────────────────────────────────────────────────────────────────────────────

\end{spacing}
\end{document}